# When Autonomous Intelligent Goodware will Fight Autonomous Intelligent Malware: A Possible Future of Cyber Defense

Dr. Paul Théron
*Thales*
*Aerospace Cyber Resilience*
*(Cyb'Air) Research Chair*
Salon de Provence, France
paul.theron@thalesgroup.com

Dr. Alexander Kott
*U.S. Army Combat*
*Capabilities Development*
*Army Research Laboratory*
Adelphi, MD USA
alexander.kott1.civ@mail.mil

*Abstract*—In the coming years, the future of military combat will include, on one hand, artificial intelligence–optimized complex command, control, communications, computers, intelligence, surveillance and reconnaissance (C4ISR) and networks and, on the other hand, autonomous intelligent Things fighting autonomous intelligent Things at a fast pace. Under this perspective, enemy forces will seek to disable or disturb our autonomous Things and our complex infrastructures and systems. Autonomy, scale and complexity in our defense systems will trigger new cyber-attack strategies, and autonomous intelligent malware (AIM) will be part of the picture. Should these cyber-attacks succeed while human operators remain unaware or unable to react fast enough due to the speed, scale or complexity of the mission, systems or attacks, missions would fail, our networks and C4ISR would be heavily disrupted, and command and control would be disabled. New cyber-defense doctrines and technologies are therefore required. Autonomous cyber defense (ACyD) is a new field of research and technology driven by the defense sector in anticipation of such threats to future military infrastructures, systems and operations. It will be implemented via swarms of autonomous intelligent cyber-defense agents (AICAs) that will fight AIM within our networks and systems. This paper presents this cyber-defense technology of the future, the current state of the art in this field and its main challenges. First, we review the rationale of the ACyD concept and its associated AICA technology. Then, we present the current research results from NATO's IST-152 Research Task Group on the AICA Reference Architecture. We then develop the 12 main technological challenges that must be resolved in the coming years, besides ethical and political issues.

I. INTRODUCTION/RATIONALE

There are at least three good reasons to advocate the concept of autonomous cyber defense (ACyD):

1. The development of autonomous weapon systems (AWS) and autonomous warfare operations.
2. The growing complexity of the information and operations technology infrastructure that will support collaborative combat.
3. The rise of autonomous intelligent malware (AIM).

First, on the battlefield of the future, "intelligent Things will fight intelligent Things" [1].

AWS are described as "weapon systems that, once activated, can select and engage targets without further human intervention" [2] or "machines acting more or less autonomously, without any direct interference from human operators" [3]. Whether on land, at sea or in the air, they are no longer fiction [4], such as drones capable of engaging in action either alone or in swarms, programmed "with a large number of alternative responses to the different challenges they may meet in performing their mission" (i.e., "with algorithms for countless human-defined courses of action to meet emerging challenges") [4].

It is difficult to predict when these technologies will become widespread [4]. However, on October 25, 2018, at the Xiangshan Forum, Zeng Yi, a senior executive at China's third largest defense company, Norinco, estimated that lethal autonomous weapon systems (LAWS) could be "common place" as early as 2025 [5]. The use of LAWS in combat already raises the highest concerns across the international community. At the UN Office at Geneva, the Convention on Certain Conventional Weapons Meeting of State Parties decided in 2013 to launch a group of experts to reflect on this matter [6], and on 12 September 2018, the European Parliament adopted a text [7] urging the EU and its Member States to work toward prohibiting LAWS.

Besides the legal and ethical aspects of the question, AWS (i.e., Intelligent Battlefield Things) will be characterized by the following:

- The autonomy of their individual decisions to engage into combat after identifying an enemy target or based on a request to engage in the context of collaborative combat operations.
- Their capacity to define or adjust their undertakings depending on the fast-paced, unpredictable, highly variable, complex circumstances of the battle theater and

- the need to coordinate with other friendly effectors in highly tactical maneuvers.
- The stealth of their moves and the likely absence of communication with a coordinating command, control, communications, computers and intelligence (C4I).

Regarding the second point of the rationale, the battlefield is a growingly complex, versatile and contested environment. Collaborative combat, in this context, is aimed at leveraging cross-systems or cross-domain synergies to gain tactical superiority.

Collaborative operations will involve and coordinate toward a common battlefield goal of manifold communication channels as well as sensing, fighting and supporting effectors operating on the ground, in the air, at sea, in space or even in cyberspace. Collaborative combat will team-up humans with the combat platforms that will support their actions [8], [1]. It will involve unmanned autonomous intelligent battlefield Things, singly or in swarms [4], because of their greater risk tolerance, for instance, in overcoming adversaries' anti-access and area-denial measures [9], thus increasing their capacity to break through enemy lines.

Programs such as the Defense Advanced Research Projects Agency's (DARPA) Collaborative Operations in Denied Environment (CODE) [10], DARPA's Gremlins [11] or the European SCAF program [12] fall under this umbrella.

Collaborative combat will then allow new, more daring, more sophisticated joint maneuvers. All of which will happen at a faster pace, especially when considering that, as said earlier, intelligent Things will be meant to fight intelligent Things [1].

Such collaborative operations will require versatile and resilient communication vectors and infrastructures. This is the very concept of an Internet of Battlefield Things [13] and the current convergence of technologies such as artificial intelligence (AI) in conjunction with software-defined networks, software-defined radio, 5G networks, etc., will create complex systems of systems constantly adapting to demands and circumstances.

The third element of the rationale is the rise of autonomous intelligent malware.

Under the previous two perspectives, enemy forces will seek to disable or disturb our autonomous battlefield Things and our complex infrastructures and systems. The enemy will need to invent new ways of attacking our systems and infrastructures. We can assume that they will employ a variety of strategies.

First, they could spread very generic, cheap malware aimed at disrupting easily accessible and fairly simple assets, as ransomware has done over the past two to three years. The extent of their spread, combined with numerous interdependencies between systems, could create widespread systemic effects and unpredictable consequences.

On another hand, enemies may seek to perform in-depth cyber-attacks, choosing to inject AIM agents into our systems. Those agents will seek pathways to targets that they will identify themselves, in context, and for which they will devise and execute attacks.

Besides targeting autonomous battlefield Things, the enemy may seek, for instance, to infiltrate the supply chain, whether via the engineering chain or maintenance contractors or staff. Pre- and post-mission connections to intelligent battlefield Things could also be eligible attack vectors.

Finally, let's assume that the cyber-attacks carried out in these contexts would be part of larger, multifaceted enemy strategies that could also include, for instance, electromagnetic attacks. Multi-vector attacks would make the defenders' situation awareness more difficult in terms of the part cyber-attacks could play in the disruption of systems and missions.

In this triple context where intelligent Things will fight other intelligent Things, where the Things might stay disconnected to remain stealthy, where collaborating platforms may be connected through the future, complex Internet of Battlefield Things in an underlying infrastructure that could be far more complex than today and where the cyber-dedicated skills pool remains limited on the battlefield, human operators in charge of monitoring cybersecurity and responding to cyber-attacks will likely be cognitively overwhelmed by the complexity, speed and scale of the events at hand.

It is also unrealistic to expect that the human warfighters residing on the platform, be they in a ground vehicle, an aircraft or a sea vessel, will have the necessary skills or time available to perform cyber-defense functions locally on the vehicle [14].

As such, current cyber-defense doctrines and technologies cannot match these future threats.

Therefore, cyber defense will need to be performed by intelligent, autonomous software agents. The agent (or multiple agents per platform) will need to stealthily monitor the networks, detect the enemy agents while remaining concealed and then destroy or degrade the enemy malware. The agents will have to do so mostly autonomously, without support or guidance from a human expert [14].

In this context, detecting, understanding and countering cyber-attacks will require a fresh approach.

## II. THE AICA REFERENCE ARCHITECTURE AS THE INITIAL STATE OF THE ART

ACyD is a new field of research and technology. It is driven by the defense sector in anticipation of threats and challenges related to future military infrastructures, systems and operations. It will be implemented via autonomous intelligent cyber-defense agents (AICAs) that, alone or in swarms, will fight AIM within our networks and systems.

The AICA Reference Architecture [14] was developed by a group of scientists from 11 countries within NATO's Science and Technology Organization / IST-152 Research Task Group (RTG) between September 2016 and September 2019.

There are several fundamental assumptions behind this technology.

AICAs will not prevent enemy malware from penetrating platforms' systems. They will be dedicated to fighting malware when it is already present within those platforms.

By being embedded within future mission-critical, safety-sensitive military effectors, systems and networks, AICAs will enable the latter to keep operating despite cyber-attacks in battlefield environments that will rely on the Internet of Battlefield Things, AI, 5G networks, autonomous effectors and vehicles, etc.

Intelligent goodware, just like future intelligent malware, will take the shape of autonomous "agents", defined as pieces of software or hardware with a processing unit capable of making smart decisions on their own about their courses of action in uncertain and adverse environments.

AICAs will have five functions, to be executed individually or collectively in swarms:

- monitor a perimeter of a host system they are to defend,
- detect signs of cyber-attacks,
- devise plans of countermeasures,
- execute tactically such plans, and
- report about their doings to human operators.

Provisions should be made for AICAs to collaborate with one agents residing on the same host system or other computers. However, in many cases, because communications might be impaired or observed by the enemy, the agents will have to avoid collaboration and operate alone [14].

AICAs will not be simple agents. Their missions, competencies, functions, architecture and technology will be a challenging construction in many ways: in terms of engineering; smart decision making, both individual and collective; their own defense against adversary malware; and the trustworthiness of the technology given ethical, societal and legal implications.

The enemy malware, its capabilities and tactics, techniques and procedures will evolve rapidly. Therefore, AICAs will need to be capable of autonomous learning [14].

In a military environment, any drawback that might be associated with the operation of AICAs will have to be balanced against the death or destruction caused by the enemy if the agent is not available [14].

The AICA Reference Architecture described in [14] is inspired from [15]. Its components are depicted in Fig. 1.

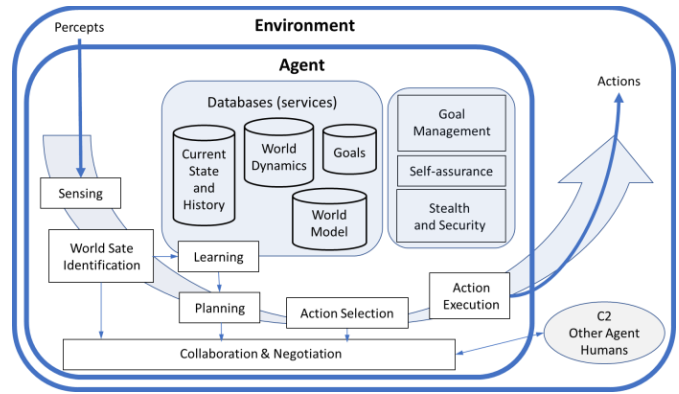

Fig. 1. AICA Reference Architecture: components of agents

AICA components include the following data services:

- World model
- World current state and history
- World dynamics knowledge
- Goals

We hypothesize that a world model is a formal descriptor of the elements it supplies to the agent's other components:

- The nominal and degraded ontology or configuration of the agent
- The nominal and degraded ontology or configuration of the system and environment (systems and threat) to defend
- The nominal and degraded ontology or configuration of cyber threats against the system and environment to defend and against the agent itself
- The nominal and degraded patterns of the world's state (agent + environment + threat). Patterns express the agent, the system or its environment's static and dynamic relations, and the concurrency of their configurations.

We hypothesize that the world current state is the evaluated distance between the world as it is and what it should be (based, for instance, on set goals or standards). Pieces of information such as the following may be required to form world state vectors describing the agent's world and that can be used by the world state identification component of AICA:

- Nominal and degraded states of reference of agents and their cohort, defended systems, their environment and connections, and threats, including the current state and the track record of past states
- Memory of cyber-defense actions and their impacts on the state of the world (current and past)
- Current data about agents and their cohort, defended systems, their environment and connections, and threats

We hypothesize that world dynamics are the following:

- An agent's behavioral rules and related expected states (nominal and degraded) in given circumstances
- Defended systems and other world objects' behavioral rules and related expected states (nominal and degraded) in given circumstances

We also hypothesize that an agent's goals are a descriptor of the rules that define the mission and the limits of the action of the agent.

Data services are not seen just as mere data repositories but as producers of processed data (i.e., "information"). It is assumed that AICAs would embark an intelligence of their own or rely on external sources to produce information, possibly in cooperation with other agent services.

The agent's data services are built in a way similar to that depicted in Fig. 2.

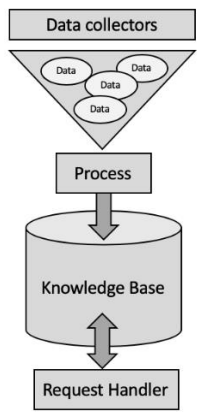

Fig. 2. General architecture of AICAs' data services

The data services will have the following characteristics [14]:

- Data collectors accept incoming data records and check their compliance to formatting and consistency rules.
- Once verified, data records are processed. Processing may be limited to mere storage instructions or the data service module may have to perform data normalization/consolidation/aggregation functions as well as exploratory data analysis and exploratory factor analysis operations.
- Data records and elaborated information can be requested by the agent's other components. In this case, the data service's request handler should be designed to check the request against validity and security rules (according to agent design options and security policies), and then data are extracted, sorted, grouped and bundled into an appropriate data container and returned to the requesting module.

On those bases, AICAs will implement five main high-level functions [14]:

1) Sensing and World State Identification
2) Planning and Action Selection
3) Collaboration and Negotiation
4) Action Execution
5) Learning

Sensing and World State Identification is the AICA high-level function that allows a cyber-defense agent to acquire data from the environment and systems in which it operates, as well as from itself, to reach an understanding of the current state of the world and, should it detect risks in it, trigger the Planning and Action Selection function.

Planning and Action Selection allows a, AICA to elaborate one to several action proposals and propose them to the action selection function, which decides the action or set of actions to execute to resolve the problematic world state pattern previously identified by World State Identification.

Action Execution allows an agent to effect the Action Selection's decision about an executable response plan (or the part of a global executable response plan assigned to the agent), monitor its execution and its effects, and provide friendly agents with the means to adjust the execution of their own part of the response plan as and when needed.

Collaboration and Negotiation enables a cyber-defense agent to 1) exchange information with other agents or a central cyber command and control (C2), or possibly a human operator, for instance, when one of the agent's functional components is not capable on its own of reaching satisfactory conclusions or usable results; and 2) negotiate with its partners the elaboration of a consolidated conclusion or result.

Learning allows an AICA to use its experience to improve progressively its efficiency with regard to all other functions.

These five high-level functions will rely upon the components of the AICA Reference Architecture described in Fig. 1.

The World State Identification component comprises the following:

- May ask the sensing component for further data if it cannot compute the current state of the environment in the agent's remit.
- Computes how good or poor the performance of previously launched plans of actions is, and if poor or inadequate, triggers the Planning component for a revision/tactical adaptation of these plans to better match the attacker's action.
- Updates, when possible/appropriate, the world current state and history, world dynamics and world model databases.

The Planning comprises the following:

- Elaborates a number of options of action (countermeasure) plans in response to the current state identified previously.

- May ask the World State Identification component for a refinement of the computation of the current state if it lacks elements to elaborate a plan of countermeasures.

The Action Selection comprises the following:

- Evaluates and ranks (in terms for instance of cost, time to deliver effects, risks, etc.) the plan options presented by the planning component.
- May ask the Planning component to refine its plan options.
- Updates the world dynamics database when it has made a clear choice of a plan and associated it with the current state found by the World State Identification component.

The Action Execution comprises the following:

- Launches the orders corresponding to the plan and sends them to the ad hoc effectors across the system defended by the agent.
- Specifies what the sensing component must monitor to supervise the execution of the action plan, and it stores those elements in the working memory to pass them on to the Sensing component.
- Updates the world dynamics database with these complementary elements of information.

The Learning component comprises the following:

- Has a generic learning mechanism that reinforces itself with experience.
- Learns on the fly from the data acquired and stored by the agent.
- Updates the database components with new elements of knowledge.
- Should trigger the ad hoc adaptations of the agent's internals to improve the latter's performance.

AICAs could be implemented in three different ways [14], and each option would entail specific choices both in terms of technology and doctrine of use:

1) A society of specialized agents: This option refers to the distributed implementation of the reference agent's functional components as a group of specialized agents, each one owning/delivering one of the functions of AICA, and the sum of the agents delivering the entire reference agent's cyber-defense capability.

2) A multi-agent system: This option refers to a swarm or cohort of fully functional agents, the architecture of which would be as in the AICA model presented previously, each one being capable of executing all AICA functions, and the swarm as a whole being supposed to deliver a collective response to a cyberattack.

3) An autonomous collaborative agent: This option refers to a fully functional agent, capable of performing on its own full cyber-defense duty on its own territory and capable, when and as needed and circumstances permitting, of communicating with other AICAs.

Should AICAs work collectively, in a swarm, for instance, in terms of their coordination, these agents could be deployed either in a centralized or decentralized manner [14].

In the decentralized approach, agents would work as master and client agents or as a distributed network of self-organizing agents. In a centralized approach, the evaluation of data and subsequent decision making could be delegated to a master agent. The master agent would control the client agents and command them to perform actions. The client agents, which would be installed on subsystem hardware could be very simple (e.g., scripts that send data and execute commands) or they could be full replicas of the designated master agent that could be activated as needed [14].

Given the accumulation of requirements and assumptions in relation to AICAs, building them is a research challenge.

III. TWELVE RESEARCH AND TECHNOLOGY CHALLENGES

In fact, AICAs are a big leap into the future of the cyber defense of military systems.

The IST-152 RTG has presented some detailed challenges that their development entails and made clear that the availability of this technology for daily military use will not happen before 10 to 15 years.

Twelve broad research challenges, not mentioning the ethical, societal and legal ones, have be identified (Fig. 3).

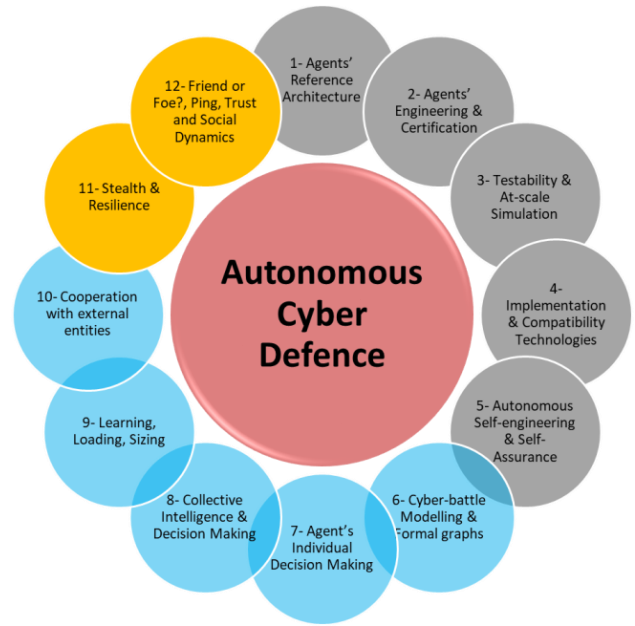

Fig. 3. Twelve research and technology challenges of ACyD

There are five engineering-related challenges:

1) Agents' reference architecture: The AICA Reference Architecture, elaborated by NATO's IST-152 RTG in

2016–2019, remains today at a very preliminary stage of specification.

2) Agents' engineering and certification: The engineering of AICAs and multi-agent systems requires multiple technical and organizational choices to be made and must be aligned with the requirements of military systems' engineering processes. Also, their certification and qualification/homologation will be complex.

3) Testability and at-scale simulation: The AICA's capacities and limitations must be systematically evaluated. To this end, simulation will be an indispensable resource. The possibility of performing such simulation at the scale of the networks and systems to defend may prove to be indispensable too.

4) Implementation and compatibility technologies: AICAs will need all the functions required to confer their autonomy, intelligence, capacity to analyze situations and make decisions, and capacity to cooperate with other agents within a multi-agent swarm or other distributed configurations. Tradeoffs between agents' functional power and the computing power and memory resource of host agents need to be explored. Also, AICAs must be compatible with host systems' cybersecurity features, otherwise the agents would be stopped or destroyed by these security features.

5) Autonomous self-engineering and self-assurance: AICAs will be embedded into host systems that will be complex and/or autonomous. They will be in operation for variable lengths of time. They will sometimes face instable conditions of communication with one another and with a central cyber C2 and human operators. It is therefore necessary to confer to them a capacity to autonomously adapt their code, algorithms or rules to respond to changes in their context of operation or to new threats and ways to combat malware.

There are five decision-making-related challenges:

1) Cyber battle modeling and formal graphs: One of the important elements that will support the decision-making process of the agent may be the "mental models", i.e., the formal representations of the host systems to defend, their resources and stakes, and also situations and moves of tactical cyber combat between goodware and malware.

2) Agents' individual decision making: Agents' decision-making is a key to their trustworthiness, but decision making is still at a very early stage of development [16]. Machine learning (ML) and reinforcement learning are regularly advocated as a pathway to the future [17], but reduce decision making to a "single loop" one-shot process. The issue with this is that in cyber battles, the adversary plans many moves and reactions to the target's response to their moves. In a tactical battle, a smart decision is one that wins the battle, not one that counters a single adversary move and risks triggering retaliation. Human decision making is smart because it builds on vigilance, vision, knowledge, experience, anticipation, wisdom, self-monitoring, deliberation, emotion and plasticity. Instance-based learning theory (IBLT) shows that five mechanisms are at play in dynamic decision making [18]: instance-based knowledge, recognition-based retrieval, adaptive strategies, necessity-based choice and feedback updates. References [19] and [20] show that for agents, making the right decision requires the integration of a variety of approaches. Decision making in action [21] suggests that the decision-making process' plasticity is an adaptive response to circumstances' characteristics and uncertainty. Allowing cyber-defense agents to make good decisions that show efficiency and value in the long run of cyber battles requires a new model of decision making. Research should thus explore the convergence of several currents of work conducted in recent years such as cognitive architectures [22] and their use for computer games [23], naturalistic decision making [24] and decision making in action [21] as they characterize the micro-cognitive processes of expert decision making, IBLT for dynamic decision making [18], agent-based modeling and simulation of cyber battles [25] and cyber-attack graphs and models [26, 27] seek to provide models of adversaries, along with game theory, AI and ML, and its current refinements.

3) Collective intelligence and decision making: To bring some form of functional superiority over enemy malware in the tactical cyber combats between goodware and malware, swarms of AICAs will need to elaborate smarter decisions based on multiple exchanges of information, mutual help or the distribution of their tasks or missions.

4) Learning, loading and sizing: Live learning within AICAs will rely on knowledge bases (such as the data services presented earlier). Such knowledge must be specified, the resource it requires, for instance, in terms of memory or processing power, must be explored. This function may prove to be too demanding in terms of resources, and offline learning and agents' preloading with downsized databases are envisaged. Here, as for the other challenges, options must be explored.

5) Cooperation with external entities: AICAs may have to or get opportunities to collaborate with a central cyber C2 system, the function of which might be to act as a central master vis-à-vis decentralized "slave" agents. They might also collaborate with human operators when they need help to raise their understanding of a situation or plan countermeasures, for instance. The protocols, the security and trustworthiness of these exchanges need to be explored.

Two challenges relate to AICAs' resilience:

1) Stealth and resilience: AICAs will be the first target of enemy malware. Their stealth will be a factor of survival for them. They will also need to be robust to attacks and able to resist degradation when attacked. Such principles will impact on the engineering of agents and need to be explored as well.

2) Friend or foe? Ping, trust and social dynamics: When AICAs "meet" or when an unknown agent meets a friendly agent or signals its presence, the latter will need to evaluate if this is a trustworthy agent with which it is possible to collaborate. Social protocols and trust evaluation need to be specified and evaluated.

IV. CONCLUSION

ACyD is a new current of research at the junction between AI (multi-agent systems, ML, deep reinforcement learning, etc.), naturalistic decision making, cybersecurity and computing science.

The NATO IST-152 RTG has made a variety of scientific publications on AICAs like the following:

- https://arxiv.org/abs/1803.10664
- https://arxiv.org/abs/1804.07646
- http://ceur-ws.org/Vol-2057
- https://arxiv.org/abs/1806.08657

It is now the duty of NATO's IST-152 RTG to define the conditions of the creation of an industry work group to further and accelerate the research and innovation efforts that ACyD requires.

Defense forces, defense industries, innovative subject-matter experts and academic research laboratories must come together to take an active part in this collective effort that current technological developments of our mission systems imply.

ACKNOWLEDGMENT

The authors thank Carol Johnson, the CCDC ARL editor, for improving the style and the grammar of the manuscript.

REFERENCES

[1] A. Kott, "Bonware to the Rescue: the Future Autonomous Cyber Defense Agents," Washington DC, Conference on Applied Machine Learning for Information Security, October 12, 2018, 018.

[2] United Nations Human Rights Council, "Report of the Special Rapporteur on extrajudicial, summary or arbitrary executions, Christof Heyns," 09 04 2013. [Online]. Available: http://www.ohchr.org/Documents/HRBodies/HRCouncil/RegularSession/Session23/A-HRC-23-47_en.pdf. [Accessed 06 10 2019].

[3] M. C. Haas, "Autonomous Weapon Systems: The Military's Smartest Toys?," The National Interest, pp. https://nationalinterest.org/feature/autonomous-weapon-systems-the-militarys-smartest-toys-11708, 20 11 2014.

[4] L. G. Dyndal, A. T. Berntsen and S. Redse-Johansen, "Autonomous military drones: no longer science fiction," NATO Review Magazine, pp. https://www.nato.int/docu/review/2017/Also-in-2017/autonomous-military-drones-no-longer-science-fiction/EN/index.htm, 28 07 2017.

[5] G. C. Allen, "Understanding China's AI Strategy. Clues to Chinese Strategic Thinking on Artificial Intelligence and National Security," Center for a New American Security, 06 02 2019. [Online]. Available: https://www.cnas.org/publications/reports/understanding-chinas-ai-strategy?utm_medium=email&utm_campaign=Understanding%20Chinas%20AI%20Strategy%20White%20Paper%20Rollout%20Media&utm_content=Understanding%20Chinas%20AI%20Strategy%20White%20Paper%20. [Accessed 07 10 20].

[6] UNOG, "Background on Lethal Autonomous Weapons Systems in the CCW," UNOG, 09 04 2013. [Online]. Available: https://www.unog.ch/80256EE600585943/(httpPages)/8FA3C2562A60FF81C1257CE600393DF6?OpenDocument. [Accessed 06 10 2019].

[7] European Parliament, "European Parliament resolution of 12 September 2018 on autonomous weapon systems (2018/2752(RSP))," European Parliament, Strasbourg, 2018.

[8] K. D. Atherton, "DARPA wants robots that humans will trust," C4ISRNET, 25 02 2019. [Online]. Available: https://www.c4isrnet.com/unmanned/2019/02/25/darpa-wants-machines-that-humans-will-trust/. [Accessed 07 10 2019].

[9] S. Savitz, I. Blickstein, P. Buryk, W. R. Button, P. DeLuca, J. Dryden, J. Mastbaum, J. Osburg, P. Padilla, A. Potter, C. C. Price, L. Thrall, S. K. Woodward, R. J. Yardley and J. M. Yurchak, "U.S. Navy Employment Options for Unmanned Surface Vehicles (USVs)," RAND Corporation, Santa Monica, California, https://www.rand.org/pubs/research_reports/RR384.html, 2013.

[10] S. Wierzbanowski, "Collaborative Operations in Denied Environment (CODE)," DARPA, [Online]. Available: https://www.darpa.mil/program/collaborative-operations-in-denied-environment. [Accessed 06 10 2019].

[11] S. Wierzbanowski, "Gremlins," DARPA, [Online]. Available: https://www.darpa.mil/program/gremlins. [Accessed 06 10 2019].

[12] Ministère des Armées, "Système de combat aérien du futur (SCAF)," French MoD, 08 02 2019. [Online]. Available: https://www.defense.gouv.fr/english/portail-defense/issues2/plf-2019/les-materiels-missions-caracteristiques-industriels/syste-me-de-combat-ae-rien-du-futur-scaf. [Accessed 07 10 2019].

[13] A. Kott, A. Swami and B. J. West, "The Internet of Battle Things," Computer, vol. 49, no. 12, pp. 70-75, 2016.

[14] A. Kott, P. Theron, M. Drašar, E. Dushku, B. LeBlanc, P. Losiewicz, A. Guarino, L. V. Mancini, A. Panico, M. Pihelgas and K. Rzadca, "Autonomous Intelligent Cyber-defense Agent (AICA) Reference Architecture, Release 2.0," US Army Research Laboratory, Adelphi, MD, 2019.

[15] S. Russell and P. Norvig, Artificial intelligence: a modern approach. 3rd ed., London (UK): Pearson, 2010.

[16] C. H. Heinl, "Artificial (Intelligent) Agents and Active Cyber Defence: Policy Implications," in 6th International Conference on Cyber Conflict, P. Brangetto, M. Maybaum and J. Stinissen, Eds., Tallinn, NATO CCD COE Publications, 2014, pp. 53-66.

[17] A. Ridley, "Machine learning for Autonomous Cyber Defense," The Next Wave, vol. 22, no. 1, pp. 7-14, 2018.

[18] C. Gonzalez, J. F. Lerch and C. Lebiere, "Instance-based learning in dynamic decision making," Cognitive Science, vol. 27, no. 2003, p. 591–635, 2003.

[19] B. Blakely and P. Theron, Decision flow-based Agent Action Planning, Prague, 18-20 October 2017: https://export.arxiv.org/pdf/1804.07646, 2018.

[20] B. LeBlanc, P. Losiewicz and S. Hourlier, A Program for effective and secure operations by Autonomous Agents and Human Operators in communications constrained tactical environments, Prague: NATO IST-152 workshop, 2017.

[21] P. Theron, "Lieutenant A and the rottweilers. A Pheno-Cognitive Analysis of a fire-fighter's experience of a critical incident and peritraumatic resilience," PhD Thesis, available at https://sites.google.com/site/cognitionresiliencetrauma, University of Glasgow, Scotland, 2014.

[22] C. Lebiere and J. R. Anderson, "A Connectionist Implementation of the ACT-R Production System," Institute of Cognitive Science University of Colorado-Boulder, 1993.

[23] P. R. Smart, T. Scutt, K. Sycara and N. R. Shadbolt, "Integrating ACT-R Cognitive Models with the Unity Game Engine," in Integrating Cognitive Architectures into Virtual Character Design, J. O. Turner, M. Nixon, U. Bernardet and S. DiPaola, Eds., Hershey, Pennsylvania, USA, IGI Global, 2016, pp. 35-64.

[24] R. Lipshitz, "Naturalistic Decision Making. Perspectives on Decision Errors," in Naturalistic Decision Making, C. E. Zsambok and G. A. Klein, Eds., Mahwah, New Jersey, Lawrence Erlbaum Associates, 1997.

[25] I. Kotenko, A. Konovalov and A. Shorov, "Agent - based simulation of cooperative defence against botnets," Concurrency and Computation. Practice and Experience, vol. 24, no. 6, pp. 573-588, 2012.

[26] S. Jajodia and S. Noel, "Advanced cyber attack modeling, analysis, and visualization," AFRL/RIGA - George Mason University, Rome NY, USA, 2010.

[27] S. Noel, E. Harley, K. H. Tam and G. Gyor, "Big-Data Architecture for Cyber Attack Graphs Representing Security Relationships in NoSQL Graph Databases," HST 2015 : IEEE Symposium on Technologies for Homeland Security, Greater Boston, Massachusetts, 2015.